\documentclass[conference]{IEEEtran}
\IEEEoverridecommandlockouts
\usepackage{cite}
\usepackage{amsmath,amssymb,amsfonts}
\usepackage{algorithmic}
\usepackage{graphicx}
\usepackage{svg}
\usepackage{glossaries}
\usepackage{cleveref}
\usepackage{textcomp}
\usepackage{url}
\usepackage{adjustbox}
\usepackage{xcolor}
\usepackage{siunitx}
\def\BibTeX{{\rm B\kern-.05em{\sc i\kern-.025em b}\kern-.08em
    T\kern-.1667em\lower.7ex\hbox{E}\kern-.125emX}}

\newacronym[longplural={Scratchpad Memories}]{SPM}{SPM}{Scratchpad Memory}
\newacronym{ACE}{ACE}{AXI Coherent Extensions}
\newacronym{AMBA}{AMBA}{Advanced Microcontroller Bus Architecture}
\newacronym{APB}{APB}{Advanced Peripheral Bus}
\newacronym{API}{API}{Application Programming Interface}
\newacronym{ASIC}{ASIC}{Application-Specific Integrated Circuit}
\newacronym{AVX}{AVX}{Advanced Vector Extension}
\newacronym{AXI}{AXI}{Advanced eXtensible Interface}
\newacronym{BLAS}{BLAS}{Basic Linear Algebra Subprograms}
\newacronym{CHI}{CHI}{Coherent Hub Interface}
\newacronym{CMOS}{CMOS}{Complementary Metal-Oxide-Semiconductor}
\newacronym{CNN}{CNN}{Convolutional Neural Network}
\newacronym[longplural={Core Complexes}]{CC}{CC}{Core Complex}
\newacronym{CPU}{CPU}{Central Processing Unit}
\newacronym{CSR}{CSR}{Control and State Register}
\newacronym{CTS}{CTS}{Clock Tree Synthesis}
\newacronym{DLP}{DLP}{Data Level Parallelism}
\newacronym{DMA}{DMA}{Direct Memory Access}
\newacronym{DRAM}{DRAM}{Dynamic Random-Access Memory}
\newacronym{DSP}{DSP}{Digital Signal Processing}
\newacronym{DUT}{DUT}{Device Under Test}
\newacronym{EW}{EW}{Element Width}
\newacronym{EEW}{EEW}{Effective Element Width}
\newacronym{ECL}{ECL}{Emitter-Coupled Logic}
\newacronym{FBB}{FBB}{Forward Body-Biasing}
\newacronym{FDSOI}{FD-SOI}{Fully Depleted Silicon-on-Insulator}
\newacronym{FMA}{FMA}{Fused Multiply-Add}
\newacronym{FPGA}{FPGA}{Field-Programmable Gate Array}
\newacronym{FP}{FP}{Floating Point}
\newacronym{FPU}{FPU}{Floating Point Unit}
\newacronym{GPGPU}{GPGPU}{General-Purpose \acrlong{GPU}}
\newacronym{GPU}{GPU}{Graphics Processing Unit}
\newacronym{HDL}{HDL}{Hardware Description Language}
\newacronym{HERO}{HERO}{Heterogeneous Embedded Research Platform}
\newacronym{HPC}{HPC}{High-Performance Computing}
\newacronym{ILP}{ILP}{Instruction Level Parallelism}
\newacronym{IoT}{IoT}{Internet-of-Things}
\newacronym{IOT}{IoT}{Internet-of-Things}
\newacronym{IPC}{IPC}{Instructions Per Cycle}
\newacronym{IPU}{IPU}{Integer Processing Unit}
\newacronym{ISA}{ISA}{Instruction Set Architecture}
\newacronym{LSB}{LSB}{Least Significant Bit}
\newacronym{LSU}{LSU}{Load/Store Unit}
\newacronym{LVT}{LVT}{low voltage threshold}
\newacronym{MIMD}{MIMD}{multiple instruction, multiple data}
\newacronym{MMU}{MMU}{Memory Management Unit}
\newacronym{MUL}{MUL}{multiplier}
\newacronym{ML}{ML}{Machine Learning}
\newacronym{NoC}{NoC}{network on chip}
\newacronym{MVL}{MVL}{maximum vector length}
\newacronym{NUMA}{NUMA}{non-uniform memory access}
\newacronym{NOC}{NoC}{Network-on-Chip}
\newacronym{PCIe}{PCIe}{Peripheral Component Interconnect Express}
\newacronym{PC}{PC}{Program Counter}
\newacronym{PE}{PE}{processing element}
\newacronym{PL}{PL}{Programmable Logic}
\newacronym{PMCA}{PMCA}{Programmable Manycore Accelerator}
\newacronym{PPA}{PPA}{Power, Performance, Area}
\newacronym{PSL}{PSL}{Power Service Layer}
\newacronym{PTE}{PTE}{page-table entry}
\newacronym{PTW}{PTW}{page-table walker}
\newacronym{PULP}{PULP}{Parallel Ultra Low Power}
\newacronym{RAW}{RAW}{read-after-write}
\newacronym{RBB}{RBB}{Reverse Body-Biasing}
\newacronym{ROB}{ROB}{Reorder Buffer}
\newacronym{RTL}{RTL}{Register Transfer Level}
\newacronym{RVT}{RVT}{Regular Voltage Threshold}
\newacronym{RoCC}{RoCC}{Rocket Custom Coprocessor Interface}
\newacronym{SCM}{SCM}{Storage Class Memory}
\newacronym{SEW}{SEW}{Standard Element Width}
\newacronym{SIMD}{SIMD}{single instruction, multiple data}
\newacronym{SIMT}{SIMT}{single instruction, multiple thread}
\newacronym{SLDU}{SLDU}{Slide Unit}
\newacronym{SLVT}{SLVT}{super-low voltage threshold}
\newacronym{SM}{SM}{Split-Mode}
\newacronym{MM}{MM}{Merge-Mode}
\newacronym[longplural={Static Random-Access Memories}]{SRAM}{SRAM}{Static Random-Access Memory}
\newacronym{SSE}{SSE}{Streaming SIMD Extension}
\newacronym{SVE}{SVE}{Scalable Vector Extension}
\newacronym{TLP}{TLP}{Thread Level Parallelism}
\newacronym{TxnID}{TxnID}{Transaction ID}
\newacronym{VAC}{VAC}{Vector Access}
\newacronym{VC}{VC}{virtual channel}
\newacronym{VCONV}{VCONV}{Vector Conversion}
\newacronym{VEX}{VEX}{Vector Execute}
\newacronym{VFU}{VFU}{vector functional unit}
\newacronym{VID}{VID}{Vector Instruction Decode}
\newacronym{VIS}{VISSUE}{Vector Instruction Issue}
\newacronym{VLIW}{VLIW}{Very Long Instruction Word}
\newacronym{VLOOP}{VLOOP}{Vector Loop}
\newacronym{VLR}{VLR}{vector length register}
\newacronym{VLSU}{VLSU}{Vector Load/Store Unit}
\newacronym{VNB}{VNB}{Von Neumann Bottleneck}
\newacronym{VRF}{VRF}{Vector Register File}
\newacronym{VT}{VT}{vector thread}
\newacronym{BW}{BW}{bandwidth}
\newacronym{MASKU}{MASKU}{Mask Unit}
\newacronym{VU0.5}{VU0.5}{Vector Unit 0.5}
\newacronym{VU1.0}{VU1.0}{Vector Unit 1.0}
\newacronym{VMFPU}{VMFPU}{Vector Multiplier/Floating Point Unit}
\newacronym{VFPU}{VFPU}{Vector Floating Point Unit}
\newacronym{VDIV}{VDIV}{Vector Divider}
\newacronym{VMUL}{VMUL}{Vector Multiplier}
\newacronym{WAR}{WAR}{write-after-read}
\newacronym{WAW}{WAW}{write-after-write}
\newacronym{DCT}{DCT}{discrete cosine transform}
\newacronym{TSV}{TSV}{through-silicon via}
\newacronym{3DIC}{3D-IC}{three-dimensional integrated circuit}
\newacronym{F2F}{F2F}{face-to-face}
\newacronym{IC}{IC}{integrated circuit}
\newacronym{C4}{C4}{controlled collapse chip connection}
\newacronym{FEOL}{FEOL}{front end of the line}
\newacronym{BEOL}{BEOL}{back end of the line}
\newacronym{SLEN}{SLEN}{striping distance}
\newacronym{VSU}{VSU}{Vector Store Unit}
\newacronym{DNN}{DNN}{Deep Neural Networks}
\newacronym{AI}{AI}{Artificial Intelligence}
\newacronym{AR}{AR}{Augmented Reality}
\newacronym{SoA}{SoA}{State-of-the-Art}
\newacronym{FD-SOI}{FD-SOI}{Fully Depleted - Silicon on Insulator}
\newacronym{RVV}{RVV}{RISC-V ``V''}
\newacronym{ALU}{ALU}{Arithmetic-Logic Unit}
\newacronym{FFT}{FFT}{Fast Fourier Transform}
\newacronym{VLDU}{VLDU}{Vector Load Unit}
\newacronym{SVC}{SVC}{Spatz Vector Core}
\newacronym{SSC}{SSC}{Snitch Scalar Core}
\newacronym{VAU}{VAU}{Vector Arithmetic Unit}

\begin{document}
\bstctlcite{IEEE:BSTcontrol}

\title{Spatzformer: An Efficient Reconfigurable Dual-Core RISC-V V Cluster for Mixed Scalar-Vector Workloads}

\newif\ifblindrev

\ifblindrev
\author{\IEEEauthorblockN{Author's list hidden for blind review}}
\else
\author{\IEEEauthorblockN{Matteo Perotti\IEEEauthorrefmark{1}\kern-.12em, Michele Raeber\IEEEauthorrefmark{1}\kern-.12em, Mattia Sinigaglia\IEEEauthorrefmark{2}\kern-.12em, Matheus Cavalcante\IEEEauthorrefmark{1}, Davide Rossi\IEEEauthorrefmark{2} and Luca Benini\IEEEauthorrefmark{1}\kern-.08em\IEEEauthorrefmark{2}
  \thanks{The first two authors contributed equally to this work.}
  \thanks{}
  \thanks{This work was supported in part through the ISOLDE (101112274) project that received funding from the HORIZON CHIPS-JU programme.}
  \thanks{}
  \thanks{\copyright 2024 IEEE. Personal use of this material is permitted. Permission from IEEE must be obtained for all other uses, in any current or future media, including reprinting/republishing this material for advertising or promotional purposes, creating new collective works, for resale or redistribution to servers or lists, or reuse of any copyrighted component of this work in other works.}}
  \IEEEauthorblockA{\IEEEauthorrefmark{1}ETH Z\"urich, Z\"urich, Switzerland, \IEEEauthorrefmark{2}Universit\`a di Bologna, Bologna, Italy}
  \IEEEauthorblockA{\{mperotti,matheus,lbenini\}@iis.ee.ethz.ch, micraebe@student.ethz.ch, \{mattia.sinigaglia5,davide.rossi\}@unibo.it}}
\fi

\maketitle

\begin{abstract}
Multi-core vector processor architectures excel in handling computationally intensive vectorizable tasks but struggle to achieve optimal resource utilization when facing sequential and control tasks that cannot be vectorized. 
This work presents Spatzformer, the first reconfigurable RISC-V V (RVV) architecture developed from a baseline open-source dual-core cluster based on Snitch scalar cores augmented with compact Spatz vector units. Spatzformer operates in two distinct modes: split mode, working as a dual-core vector architecture to handle vectorizable tasks concurrently, and merge mode, where two vector units are driven by a single scalar core, allowing the remaining scalar core to handle non-vectorizable control tasks. 
We implement Spatzformer in a 12-nm technology node and characterize the cost of the added architectural reconfigurability. We show that merge mode accelerates mixed scalar-vector kernels by up to $1.8 \times$ compared to split mode. Moreover, it accelerates the vector kernels that require fine-grained synchronization (such as FFT) by up to 20\% with respect to the baseline. 
The reconfigurability features do not degrade the architecture's maximum frequency (\qty{1.2}{\giga\hertz}, TT, \qty{0.8}{\volt}, \qty{25}{\celsius}) and have a negligible area impact (+1.4\%), with a worst-case energy efficiency drop of only 7\% with respect to the non-reconfigurable baseline.
\end{abstract}


\section{Introduction}
Vector processing represents a compelling solution to the ever-growing demand for computational power and energy efficiency. A single vector instruction operates on multiple data elements, reducing the energy overhead of instruction fetch and dispatch. Furthermore, the large \gls{VRF} of vector architectures enhances the data reuse of computationally intensive applications \cite{cavalcante2023spatz}. 

The flexibility of multi-core vector computing systems has been praised as a way to tackle the increasing complexity and heterogeneity of typical workloads. This is testified by the numerous recent solutions from industry~\cite{9355239, SiFiveX280, AndesNX27V} and academia~\cite{cavalcante2023spatz}, boosted by the recent freezing of the open-source \gls{RVV} Vector \gls{ISA}. These architectures excel in task parallelization, especially with regular workloads, making them well-suited for a wide range of applications, from high-performance computing to embedded systems.

Despite their adaptability and growing adoption, multi-core vector architectures struggle to achieve high resource utilization when confronted with workloads characterized by extreme diversity, such as the simultaneous processing of parallel workloads and heavily sequential control tasks, e.g., with critical applications like autonomous driving and radar processing.
In such cases, the architecture must either serialize the execution of vector and scalar kernels or allocate one of the vector cores to handle the scalar task. This leads to suboptimal use of the vector computational resources.

In a nutshell, in this work: 1) we present Spatzformer, the first reconfigurable multi-core vector architecture based on RVV. Spatzformer is based on the open-source dual-core Spatz vector processor architecture \cite{cavalcante2023spatz} and can be reconfigured at runtime in two operational modes to adapt to heterogeneous parallel workloads, which include control or serial tasks that need to be run alongside vector kernels; 2) we implement Spatzformer in a 12-nm technology and characterize the cost of the added reconfigurability in terms of \gls{PPA}, including a performance and energy-efficiency analysis on six kernels from different domains; 3) we analyze the performance gain of Spatzformer over the baseline cluster when executing mixed scalar-vector workloads.

Our reconfigurable architecture aims at improving the flexibility of traditional chip-oriented multi-core vector systems, providing a versatile and efficient solution tailored for heterogeneous computational and control applications where energy efficiency and area constraints are paramount.

With our work, we study the cost of adding lightweight reconfigurability to a highly optimized chip-oriented vector architecture and analyze its performance benefits and trade-offs. This differs from other similar works, such as \cite{reconfigDualCore} and \cite{elasticOccamy}, that do not focus on the \gls{PPA} cost of the reconfigurability feature itself.

\section{Spatzformer Architecture}\label{sec:architecture}
\begin{figure}[t]
  \centering
  \includegraphics[width=\linewidth]{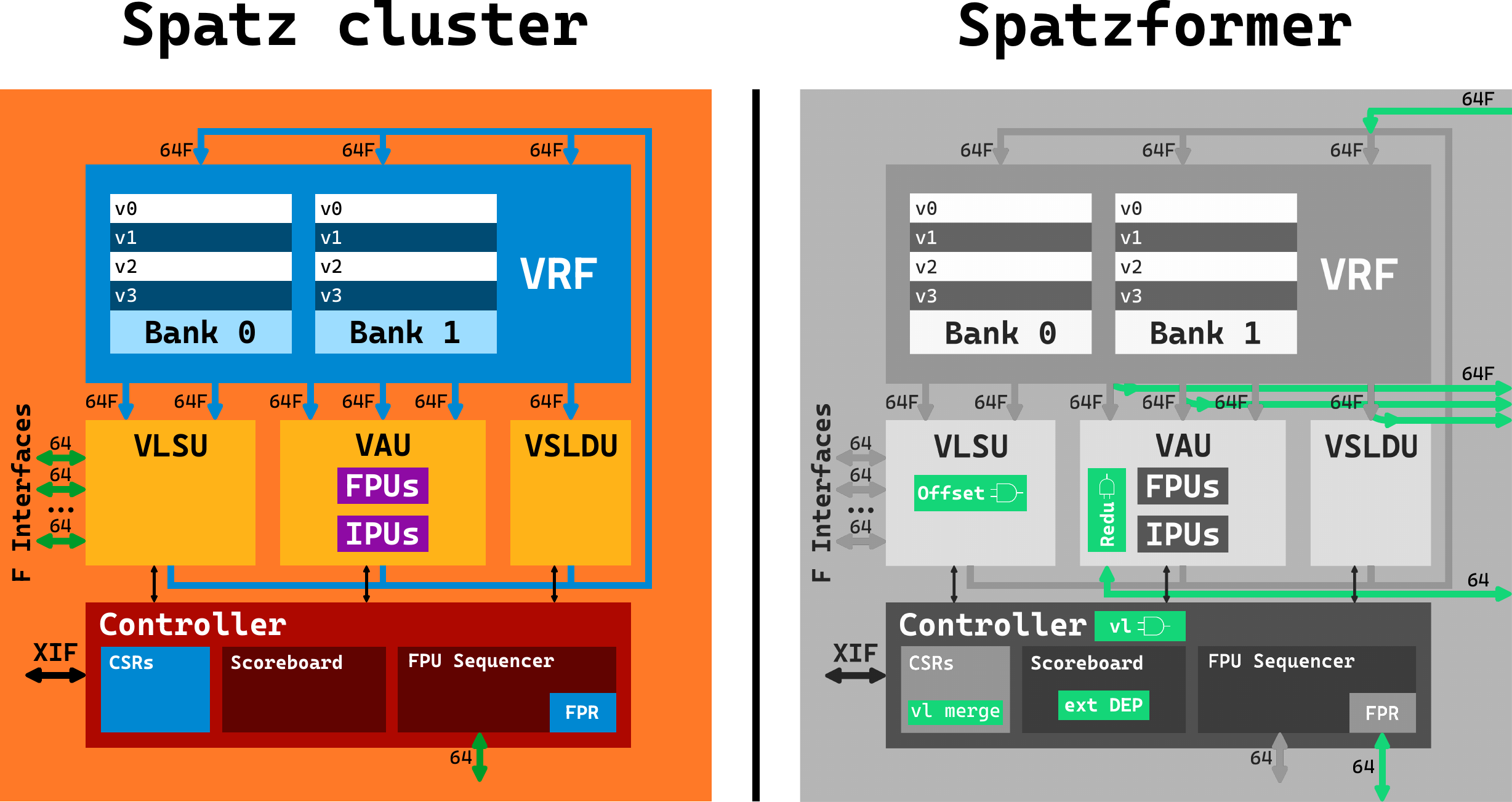}
  \caption{Baseline Architecture and Spatzformer's reconfigurability logic.}
  \label{fig:baseline_architecture}
\end{figure}

Spatzformer can be configured at runtime by the programmer in two operational modes. In \gls{SM}, the architecture is composed of two scalar cores, each of which is coupled with its own vector accelerator. In \gls{MM}, one scalar core drives both vector accelerators, while the remaining scalar core can process sequential control tasks without slowing down the vector execution. The operational mode can also change at runtime.
\Cref{fig:baseline_architecture} shows the architecture of the baseline Spatz cluster and Spatzformer's microarchitectural modifications that enable runtime reconfigurability.

\section{Results}\label{sec:results}
To assess the implementation cost of \gls{RVV} reconfigurability, we synthesize and place-and-route the baseline architecture and Spatzformer in an advanced 12-nm technology node. 

\begin{figure}[t]
  \centering
  \includegraphics[width=\linewidth]{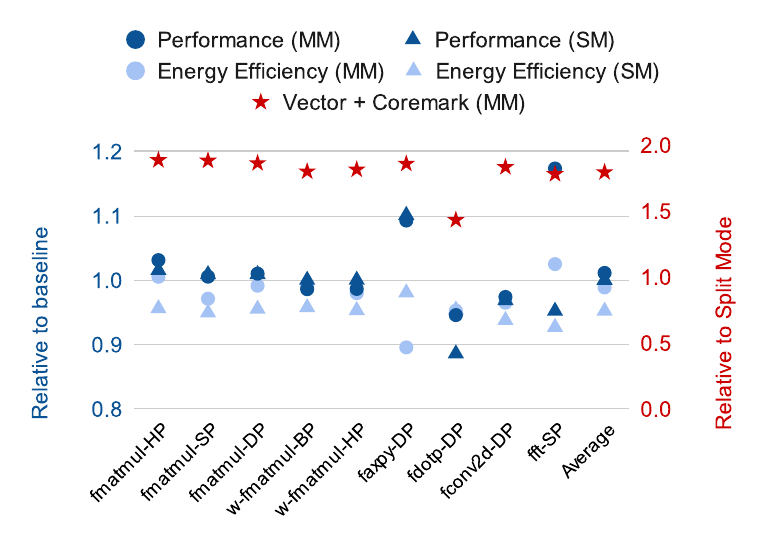}
  \caption{Performance and Energy Efficiency comparison of Spatz cluster (baseline) and Spatzformer in split and merge mode (left axis), and merge mode speed up of mixed scalar-vector workload over split mode (right axis).}
  \label{fig:res_plot}
\end{figure}

To evaluate how reconfigurability affects the architecture's performance and energy efficiency, we simulate six vector kernels with various degrees of data reuse and arithmetic intensity in \gls{SM} and \gls{MM} against the split-mode-only baseline. The main results are summarized in \Cref{fig:res_plot} (left axis).

\textbf{Area}: 
Spatzformer requires only an area overhead of 55 kGE (+1.4\%), as opposed to adding a dedicated scalar core to achieve comparable parallelization of scalar-vector tasks, which would require an area inflation of at least +6\% (more than $4\times$ larger).

\textbf{Performance and Energy Efficiency}: The added reconfigurability feature does not hurt the maximum frequency of the Spatz cluster (950 MHz, SS, $0.72\text{V}$, $125^\circ\text{C}$).

The average performance across all vector benchmarks shows that Spatzformer is as fast as the baseline when executing vector kernels in \gls{SM} and can outperform it in \gls{MM}, with only a slight average energy efficiency loss of 5\% (\gls{SM}) and 1\% (\gls{MM}) from the baseline architecture.

Notably, \gls{MM} \texttt{fft} outperforms \gls{SM} \texttt{fft} by more than 20\% and shows a 2.5\% higher energy efficiency by reducing the synchronization overhead of the multi-core architecture.  

Spatzformer's power consumption in \gls{MM} is negatively impacted by the added hardware to support reconfigurability and positively affected by the savings that come from \gls{MM}. Indeed, \gls{MM} reduces the energy related to the instruction fetch from memory thanks to the higher vector length on which instructions are amortized.

\textbf{Mixed scalar-vector workload}:
Finally, we show how \gls{MM} improves scalar-vector mixed workload performance by running in \gls{SM} and \gls{MM} our vector kernels in parallel with CoreMark \cite{coremark}, an industry-grade benchmark that simulates common workload executed by scalar cores. 

The \gls{MM} Spatzformer boosts the baseline performance by up to almost $2\times$ in the best case, with an average of 1.8$\times$, showing that \gls{MM}-Spatzformer can hide the latency of simple control tasks that use the memory without significant performance drops on the vector kernels.

\section{Conclusion}\label{sec:conclusion}
We presented Spatzformer, a reconfigurable \gls{RVV} dual-core architecture, and assessed the cost of the added reconfigurability features in terms of \gls{PPA} and energy efficiency on a 12-nm technology node implementation executing kernels from \gls{ML}, \gls{DSP}, and Linear Algebra.

Spatzformer implements reconfigurability with a negligible 1.4\% area overhead and a slight average 5\% drop in energy efficiency when executing vector kernels in split mode and no maximum frequency degradation.
Merge mode helps speed up mixed scalar-vector workloads (1.8$\times$ on average) and tasks like FFT (+20\%) by freeing up a scalar core and reducing the multi-core synchronization overhead. Split mode, instead, can be used to work on different tasks in parallel or flexibly exploit a second dimension of parallelization on a single application.

\bibliographystyle{IEEEtran}
\bibliography{ieeetran,main}

\begin{thebibliography}{1}
\providecommand{\url}[1]{#1}
\csname url@samestyle\endcsname
\providecommand{\newblock}{\relax}
\providecommand{\bibinfo}[2]{#2}
\providecommand{\BIBentrySTDinterwordspacing}{\spaceskip=0pt\relax}
\providecommand{\BIBentryALTinterwordstretchfactor}{4}
\providecommand{\BIBentryALTinterwordspacing}{\spaceskip=\fontdimen2\font plus
\BIBentryALTinterwordstretchfactor\fontdimen3\font minus \fontdimen4\font\relax}
\providecommand{\BIBforeignlanguage}[2]{{%
\expandafter\ifx\csname l@#1\endcsname\relax
\typeout{** WARNING: IEEEtran.bst: No hyphenation pattern has been}%
\typeout{** loaded for the language `#1'. Using the pattern for}%
\typeout{** the default language instead.}%
\else
\language=\csname l@#1\endcsname
\fi
#2}}
\providecommand{\BIBdecl}{\relax}
\BIBdecl

\bibitem{cavalcante2023spatz}
M.~Cavalcante, M.~Perotti, S.~Riedel, and L.~Benini, ``Spatz: Clustering compact {RISC-V}-based vector units to maximize computing efficiency,'' arXiv:2309.10137 [cs.AR], 2023.

\bibitem{9355239}
M.~Sato \emph{et~al.}, ``Co-design for {A64FX} manycore processor and `{Fugaku}','' in \emph{SC20}, 2020, pp. 1--15.

\bibitem{SiFiveX280}
\BIBentryALTinterwordspacing
\emph{{SiFive} Intelligence {X280}}, {SiFive Corp.}, San Mateo, CA, USA, 2022, revision 21G3. [Online]. Available: \url{https://sifive.cdn.prismic.io/sifive/62e0df53-be02-4b50-b211-aa55b7042fc8_x280-datasheet-21G3.pdf}
\BIBentrySTDinterwordspacing

\bibitem{AndesNX27V}
\BIBentryALTinterwordspacing
\emph{{AndesCore} {NX27V}}, {Andes Technology}, Hsinchu City, Taiwan, 2020. [Online]. Available: \url{http://www.andestech.com/en/products-solutions/andescore-processors/riscv-nx27v}
\BIBentrySTDinterwordspacing

\bibitem{reconfigDualCore}
S.~F. Beldianu and S.~G. Ziavras, ``Multicore-based vector coprocessor sharing for performance and energy gains,'' \emph{ACM Trans. Embed. Comput. Syst.}, vol.~13, no.~2, sep 2013.

\bibitem{elasticOccamy}
Z.~Zhang \emph{et~al.}, ``Occamy: Elastically sharing a {SIMD} co-processor across multiple {CPU} cores,'' ser. ASPLOS 2023.\hskip 1em plus 0.5em minus 0.4em\relax New York, NY, USA: Association for Computing Machinery, 2023, p. 483–497.

\bibitem{coremark}
\BIBentryALTinterwordspacing
EEMBC, ``{CoreMark} benchmark.'' [Online]. Available: \url{https://github.com/eembc/coremark}
\BIBentrySTDinterwordspacing

\end{thebibliography}

\end{document}